\documentclass[nohyper,12pt,letterpaper]{JHEP}

\newfont{\frak}{eufm10 scaled 1200}

\newfont{\Bbb}{msbm10 scaled 1200}     
\newcommand{\mathbb}[1]{\mbox{\Bbb #1}}
\DeclareSymbolFont{AMSa}{U}{msa}{m}{n}
\DeclareSymbolFont{AMSb}{U}{msb}{m}{n}
\let\Box\relax
\DeclareMathSymbol{\Box}{\mathord}{AMSa}{"03}

\def\CO{{\cal O}}
\def\CL{{\cal L}}
\def\half{{1\over 2}}

\def \eqn#1#2{\begin{equation}#2\label{#1}\end{equation}}

\title{Time-Varying $\alpha$ and Particle Physics}

\author{T. Banks\\
Department of Physics and Astronomy\\ Rutgers University NHETC,
Piscataway, NY
 \\
Department of Physics\\ U.C. Santa Cruz/SCIPP, Santa Cruz, CA\\
E-mail: \email{banks@physics.rutgers.edu}}

\author{M. Dine\\
Department of Physics\\ U.C. Santa Cruz/SCIPP, Santa Cruz, CA\\
E-mail: \email{dine@scipp.ucsc.edu}}

\author{Michael R. Douglas\\
Department of Physics and Astronomy\\ Rutgers University NHETC,
Piscataway, NY
\\
I.H.E.S., Bures-sur-Yvette 91440 France \\
E-mail: \email{mrd@physics.rutgers.edu}}

\abstract{We argue that models in which an observable variation of
the fine structure constant is explained by motion of a cosmic
scalar field, are not stable under renormalization, and require
massive fine tuning that cannot be explained by any known
mechanism. }


\received{???????? ?st, 2000} \accepted{???????? ?th, 1998}
\preprint{\hepth{0112059}\\RUNHETC-2001-\\SCIPP-\\}

\begin{document}




\section{\bf Introduction}

Recent observations of distant quasars have revived
the suggestion that the fine structure constant varies over
cosmological time scales.  The observations of \cite{quasar} suggest
a variation of
${\delta\alpha \over \alpha} \sim 10^{-4}$ over the time period
since emission of the quasar light.

Within both the contexts of effective
field theory and M-theory, it is natural to model the change of
the fine structure constant by coupling a dynamical scalar field
$\phi$ to the photon kinetic term in the low energy effective
action.

\eqn{deltaalpha}{\delta e^{-2} = \epsilon {\phi\over M}
F_{\mu\nu}^2}

One can imagine two broad classes of scenarios to explain the putative
variation of $\alpha$.  Usually, a slow, secular variation of $\phi$
is postulated.  Another possibility is that there was a first order
phase transition between the time at which the quasar light was
emitted and the present.  In this note we will argue that any such
time variation of $\alpha$ raises difficulties.  In particular, such a
large variation of $\alpha$ can only be compatible with basic
principles of quantum field theory if there is an extraordinary degree
of fine tuning of many parameters of the underlying theory.\footnote{
This conclusion has also been reached by N. Kaloper and L. Susskind
\cite{KS}.}

Let us first state the basic argument, and later show how this comes
out of concrete calculations.  It is that the vacuum energy, as
computed in the standard model, or in more general low energy
effective field theory, must depend on $\alpha$, even taking into
account our ignorance about the resolution of the cosmological
constant problem.  This dependence can be estimated in various ways,
and combining these estimates with the criterion that the new
contributions do not dominate the energy densities which appear in
standard cosmology, leads to fantastically tight bounds.

In generic quantum field theory, varying an arbitrary dimensionless
coupling such as $\alpha$ will lead to a variation in the vacuum energy
$V$ controlled by the cutoff scale $\Lambda$,
\eqn{vacen}{
\delta V = c~ \delta \alpha~ \Lambda^4
}
with $c$ an $\CO(1)$ constant.  For example, in QED this is
$<F_{\mu\nu}F^{\mu\nu}>$ and in perturbation theory $c=1/\pi+\ldots$.
Perturbative corrections and the effects of
other matter in the standard model will modify
this result, but again lead to $c\sim \CO(1)$.

The estimate, \ref{vacen} might well be an overestimate.  If,
throughout the relevant cosmic history, the field $\phi$ has been
near its minimum, the vacuum energy will be of order $\delta
\alpha^2$,
\eqn{secondorder}{\delta V = C \alpha ({\delta \alpha \over \alpha})^2 \Lambda^4.}
In models of quintessence or in which the change in $\alpha$ is
due to a phase transition, the first order estimate is likely to
be correct.

Let us be extremely conservative and take $\Lambda$ to be the QCD scale
$\sim 100~ {\rm MeV}$.  Then
\eqn{changevac}{
\delta V = ({\delta \alpha \over \alpha}) 10^{29} {\rm (eV)^4}
}
However, the matter dominated era of conventional cosmology has
a maximum energy density of order $10^4 {\rm eV}^4$.  Thus, some of what
is supposed to be the matter dominated era, and in particular the period
when the quasar light was emitted, was instead dominated by a large
scalar field potential energy.  This changes classical cosmology
completely, and is ruled out by observation.   At the earliest stages of
galaxy formation, the energy density was of order $10^{-8} {\rm eV}^4$.
This argument leads to the bound, if the variation is first order,
\eqn{alphabound}{
\left|{\delta\alpha \over \alpha}\right| < 10^{-37} .
}
The bound is significantly weaker if the change is second order,
of order $10^{-18}$, but still this estimate is
many orders of magnitude smaller than the variation inferred from
the quasar observations.  It is necessary to suppress not only the
quadratic terms in the $\phi$ potential, but terms up to very high
(roughly eighth or ninth)
order, to accommodate a variation of order $10^{-4}$.

As we noted, these estimates followed from an extremely conservative choice of
$\Lambda$.  Most of us believe that physics is well described by field
theory at scales below some higher $\Lambda$, possibly as high as the
Planck scale $M_P$ (this would lead to $\left|{\delta\alpha/
\alpha}\right| < 10^{-104}$).  There is no evidence that this general
behavior depends on the form of the theory above the cutoff scale.  In
particular, corrections of this form invariably arise in string
theory, in instances where one can do the calculation.

The situation is not quite so extreme in supersymmetric theories,
which do generically
cancel the vacuum energy.  Of course in the real world supersymmetry is
broken, at some scale $M_{SUSY}$.  The generic estimate in this situation
is
\eqn{susyvacen}{
\delta V = c~ \delta \alpha ~ \Lambda^2 M_{SUSY}^2 .
}
Supersymmetry generically cancels leading order divergences, but does
not do better than that.  In the most optimistic scenario,
a particular
supersymmetric model might indeed cancel all divergences, leading to
$\delta V \propto M_{SUSY}^4$.  Since experiment constrains
$M_{SUSY} > 100 {\rm GeV}$, even if
\begin{equation}
\delta V = (\delta \alpha)^2 M_{SUSY}^4
\end{equation}
we obtain ${\delta \alpha \over \alpha} < 10^{-24}$.

Note that this argument would apply both to a continuous variation
of $\alpha$ (we will study this in more detail below), but also to
more drastic modifications of the physics between early and late
times, such as a first order phase transition.  Indeed, on general grounds
one expects the vacuum energy to decrease with time; in this situation
a phase transition can only increase the variation of the energy.

Let us now address a possible objection to this argument.  It is that
normally, we regard terms such as (\ref{vacen}) as part of the
cosmological constant, and imagine that they are subtracted off by
whatever mysterious agency resolves the cosmological constant problem.

However, if $\alpha$ varies, general physical principles force us to
treat it effectively as a scalar field.  Then, we should not do this
subtraction for all values of $\alpha$ but only for a particular value:
its asymptotic value, or perhaps its value at the current era. There
are many reasons to believe this:

\begin{itemize}

\item Subtracting the cosmological constant in low energy effective
field theory means fine tuning the coefficient of a single relevant
operator. Subtracting the vacuum energy for all values of a scalar field
means fine tuning the coefficients of an infinite number of relevant and
irrelevant operators.

\item All theories of inflation and quintessence assume that the values
of the potential away from the minimum of a scalar field are nonzero.
Indeed, if we imagined the solution of the cosmological constant problem
actually subtracted away the potential of all scalar fields, we would
not even be able to implement the Higgs mechanism.

\item Specific proposals {\it e.g.} \cite{tbfolly} for explaining the
mystery of the cosmological constant, involve fine tuning only the
minimum value of the potential.

\end{itemize}

Thus, this objection does not invalidate our conclusion, that
observable variations of the fine structure constant are highly
unnatural.

To complete the discussion, we now study the cosmology of a field
$\phi$ as postulated in (\ref{deltaalpha}), to see what type of
violations of naturalness are required to evade the bound.
Besides the coupling (\ref{deltaalpha}),
we take $\phi$ to be governed by the Lagrangian
\eqn{philagrangian}{
\CL = \half(\partial\phi)^2 - V(\phi).
}
The potential energy $V(\phi)$ will be modeled by the generic form
$\mu^4 f(\phi /M)$.  If we neglect Hubble friction, the natural time
scale for motion of this field is $M\over \mu^2$. For plausible values
of the microphysical parameters $\mu$ and $M$, this motion will be too
rapid to fit the slow time variation of $\alpha$ .  We therefore
assume that, until very recently, the motion has been friction
dominated.  This requires $M H/\mu^2 >> 1$ and  $M >> M_P/48\sqrt{\pi}$.
Then

\eqn{var}{\delta \alpha^{-1} = {4\pi \epsilon \mu^{4} \over 3 M^2}
\int {dt f^{\prime} ({\phi\over M})\over H}}

The integral is over a period in the matter dominated era of the
universe.  During this period, $\phi / M$ changes only very
little, while $f$ is a smooth function of order one.  Thus

\eqn{varb}{\delta \alpha^{-1} \sim {\epsilon \mu^{4} \over
M^2}{M_P^2 \over \rho_{NOW}}}

The energy density $\mu^4$ can at most be of order $\rho_{NOW}$,
since otherwise we would have seen inflation rather than matter
domination and the calculation would not be self consistent.  If
it is of order $\rho_{NOW}$ then $\phi$ is a form of quintessence
\cite{rhatra,hill}.  Thus we have:

\eqn{varc}{\delta \alpha^{-1} \leq {\epsilon M_P^2 \over M^2}}

At this point, one might conclude that $\CO(1)$ values for $\epsilon$
and $M/M_P$ could lead to the desired result.  However, we still need
to consider the effect of $\epsilon$ on the vacuum energy.  This can
be estimated along the lines discussed earlier.  We assume a supersymmetric
theory, to obtain
\eqn{potchang}{
\delta V = (\Lambda M_{SUSY})^2 g(\epsilon \phi /M)
}
in terms of a smooth $\CO(1)$ function $g$.  We furthermore assume
$\Lambda\ge M_{SUSY}$ and then set $\Lambda=M_{SUSY}$ to obtain a
conservative bound.

In order that this interaction does not
change our estimate of the size of the potential, we must insist that
\eqn{epsbnd}{
\epsilon \leq {\sqrt{\rho_{NOW}} \over M_{SUSY}^2,}
}
and this gives a stringent bound on the time variation of the fine
structure constant:
\eqn{vard}{
{\delta \alpha \over \alpha} \leq \alpha
{\rho_{NOW}^{1/2} M_P^2 \over M^2 M_{SUSY}^2} \sim 10^{-28} ({M_P
\over M})^2 .
}
Thus taking ${\delta \alpha} \sim 10^{-4}\alpha$
is inconsistent with our initial assumptions.  That is,
a field with $M < 10^7$ GeV and $\mu^4 \sim \rho_{NOW}$ would not
undergo friction dominated motion during any part of the matter
dominated era of the universe.  Restoring consistency by requiring
$M>M_P$ brings us back to the bound (\ref{alphabound}).

The essential point is the same as in the first argument, that for
such a variation to be consistent with cosmology requires fine tuning
of the potential over a range of parameters.  Indeed, if we want
$V(\phi) \sim \rho_{NOW}$ over the entire range
$0\le\delta\alpha\le 10^{-4}\alpha$, we must fine tune away contributions
from (roughly) the first ten coefficients of the Taylor expansion of
the function $g$ in (\ref{potchang}).

A potential loophole in the argument as we just formulated it is that
we assumed the coupling (\ref{deltaalpha}), but this precludes the
interesting case that $V(\phi)$ vanishes at infinity, and that the
scalar is evolving towards arbitrarily large values.  Many models of
quintessence assume such a potential.

This point can be dealt with by generalizing the Lagrangian to
\eqn{philagrangiantwo}{
\CL = \half g(\phi)(\partial\phi)^2 - V(\phi).
}
Here $g(\phi)$ is an arbitrary positive function which
might be thought of as related to wave function renormalization, or better as
a metric on the configuration space.
Within this family of Lagrangians, we can perform arbitrary
field redefinitions $\phi \rightarrow \phi'(\phi)$.

In fact we can use this freedom to redefine $\phi$ so that the
coupling (\ref{deltaalpha}) is exact; in other words define $\phi$ by
the relation $\epsilon\phi/M = \delta\alpha^{-1}$, the variation
$\delta\alpha$ around its late time asymptotic value.
The possibility
we missed of $\phi$ going to infinity becomes, after the field redefinition,
the possibility that the integral $\int d\phi g(\phi)^{1/2}$, the invariant
measure of distance in field space, could diverge as $\phi\rightarrow 0$.
\footnote{
We are not being perfectly general yet as we are assuming that the
original relation between $\phi$ and $\delta\alpha$ was one-to-one.
One can generalize further, but this does not lead to further illumination.
}

Under our previous assumptions, the slow roll analysis goes through
in the same way, with the equation of motion
$$
\dot\phi = - g^{-1}(\phi) V'(\phi)
$$
and corresponding modifications to (\ref{var}) and other equations.
However, the main point does not require any detailed analysis.
It is that the new possibility, in this language that $g(\phi)$ diverges
as $\phi\rightarrow 0$, would lead to a slowing down of $\dot\alpha$,
but does not affect the essential point as seen in the first argument.

In general, the limit $\phi>>\Lambda$, which might be thought to be
problematic in effective field theory, in many cases is not.  Such
field redefinitions can also be used to clarify other limits; for
example $M>>M_P$, etc.

All this is not to say that scalars with potentials $V(\phi) \sim
\rho_{NOW}$ are impossible.  This still requires an extreme fine
tuning, of course, but if we place suitable bounds on $\epsilon$,
leading to unobservably small $\delta\alpha/\alpha$, one can argue
that one has done only two fine tunings, of $\mu/M_P$ and of
$\epsilon$.  The most natural way to accomplish the second of these
would be to simply postulate a symmetry enforcing $\epsilon=0$.

As far as we know, the only possible explanation of such small numbers
would be an axion-like shift symmetry for the scalar field
$\phi$. General arguments rule out global continuous symmetries in
theories including gravity.  However, there are axion like fields that
arise from higher dimensional antisymmetric tensor gauge fields. A
generic context in which one might expect small axion potentials to be
generated, is that of brane world models, like that of Horava and
Witten \cite{horwit}.  There, axions arise as would be gauge modes of
bulk gauge fields, and potentials are generated by interaction with
the boundary.  In some Horava-Witten models for the real world, one
can obtain axions whose potentials are generated by weak interaction
instantons\cite{bd} .  This can give rise to very small numbers for
couplings that would be order one by dimensional analysis.

This reinforces the conclusion of \cite{carroll} that axions are the
only known model of quintessence that might be consistent with the
naturalness constraints of quantum field theory.  Note however that
\cite{carroll}\cite{choi} and \cite{bdark} all point out various
problems with the idea of axions as quintessence.

Our overall conclusion is that we do not have {\it any} field
theoretically natural explanations for a variation of the fine
structure constant as large as would be required to explain the
observations of \cite{quasar}.  If these observations are
confirmed, one will have to invent some very exotic physics to
explain them.

\medskip

{\it Note added:}
In this note, we have focused on the variation of the fine
structure constant, since this is the manner in which the result
of \cite{quasar} is presented.  However, in any model in which the
variation of a field leads to variation of the fine structure
constant, it is likely to lead to variation of the other constants
of nature, including the electron mass and the QCD coupling, as
well as (if nature is supersymmetric)
the scale of supersymmetry breaking.
Arguments suggesting correlated variations of several couplings
have recently been made in \cite{cflss}.

Following the philosophy we are advocating,
and again assuming that we are near a minimum of the effective potential,
general arguments about minima of functions suggest that any
correlated variation of couplings will yield a similar bound.
In particular, the limits on the fractional
variation of the electron mass and the QCD coupling will be even more severe
than those we have discussed here.

\section{Acknowledgments}

The research of M.R.D. and T.B. was supported in part by DOE grant
number DE-FG02-96ER40959, and the research of M.D. and T.B. was
supported in part by DOE grant number DE-FG03-92ER40689.

%


\end{document}